# Magnetic Helicity Signs and Flaring Propensity: Comparing Force-free Parameter with the Helicity signs of Hα Filaments and X-ray Sigmoids

V. Aparna [ID],[1,2,3] Manolis K. Georgoulis [ID],[4,5] and Petrus C. Martens [ID][3]

[1]*Lockheed Martin Solar and Astrophysics Laboratory, 3251 Hanover Street, Building 203, Palo Alto, CA 94304, USA*
[2]*Bay Area Environmental Research Institute, 625 2nd Street, Suite 209, Petaluma, CA 94952, USA*
[3]*Department of Physics & Astronomy, Georgia State University, 625, 25 Park Place, Atlanta, GA 30303*
[4]*Space Exploration Sector, Johns Hopkins Applied Physics Lab, Laurel, MD 20723 USA*
[5]*Research Center for Astronomy and Applied Mathematics, Academy of Athens, 115 27 Athens, Greece*

## ABSTRACT

Sigmoids produce strong eruptive events. Earlier studies have shown that the ICME axial magnetic field Bz can be predicted with some credibility by observing the corresponding filament or the polarity inversion line in the region of eruption and deriving the magnetic field direction from that. Sigmoids are coronal structures often associated with filaments in the sigmoidal region. In this study, firstly we compare filament chirality with sigmoid handedness to observe their correlation. Secondly, we perform non-linear force-free approximations of the coronal magnetic connectivity using photospheric vector magnetograms underneath sigmoids to obtain a weighted-average value of the force-free parameter and to correlate it with filament chirality and the observed coronal sigmoid handedness. Importantly, we find that the sigmoids and their filament counterparts do not always have the same helicity signs. Production of eruptive events by regions that do not have the same signs of helicities is ∼3.5 times higher than when they do. A case study of magnetic energy/ helicity evolution in NOAA AR 12473 is also presented.

## 1. INTRODUCTION

Monitoring coronal mass ejections (CMEs) and solar flares occurring on the Sun is extremely important for various reasons - humans venturing into becoming a multiplanetary species, protection of astronauts onboard the international space station and elsewhere from high-energy particles, oil pipelines on the ground from induction of magnetic fields and overcharging of satellite electronics (Lam et al. 2012). The daily lives of human beings depend on the above facilities, for which uninterrupted operations should be secured, and must be well protected for their uninterrupted functioning. To this end, there have been several forecasting efforts that utilize remote sensing and in-situ data collected by satellites to predict extreme space weather events, for example - Kubo et al. (2017); Temmer (2021); Whitman et al. (2022). Major flares and fast CMEs stem exclusively from solar active regions while the quiet Sun is responsible for slower CMEs, with the fastest of them close to the speed of fast solar wind. In this study, we focus on the coronal sources of strong solar eruptive events, namely X-ray sigmoids, and their relation with their counterparts in chromospheric wavelengths - namely solar filaments. Our objective here is to determine how sigmoids and filaments can be used in conjunction for the prediction of intense space weather events.

Recently, Aparna & Martens (2020) and Mundra, Aparna, & Martens (2021) have shown that solar filaments can be effectively used for predicting geo-effective events by determining the direction of the filament axis. They use small scale structures such as barbs and fibrils to determine the filament chirality which combined with the polarity information, gives the direction of the axis. Sigmoidal active regions are known for producing a large number of eruptive events (Canfield et al. 1999; Sterling 2000; Savcheva et al. 2014; Kawabata et al. 2018) that can cause a significant space

Corresponding author: V. Aparna
aparna@baeri.org

weather impact. Recently, in a statistical analysis of eruptive event occurrence in sigmoids, Kawabata et al. (2018) observed that only a very small number (∼9%) of CMEs were not associated with a sigmoidal region and that "sigmoid structures have a stronger dependence on CME occurrence than large X-ray flares". Sigmoids are large X-ray structures that are easily detected on the Sun and can be programmed to be identified for their shapes - which are representative of their magnetic helicity, via computer vision, which may then be used for predicting space weather events, if they have a one to one correlation with the filaments. Our work is motivated by the need of the hour to have automated algorithms to effectively predict space weather events. To accomplish this, we first investigate if the sigmoids and filaments have a one to one correlation in terms of their magnetic structure. We then provide a quantitative analysis of the helicities of the sigmoids and filaments. In the following subsections, we first give an overview of the sigmoid and filament observations along with some details of their magnetic structures.

## 1.1. *Sigmoids*

Sigmoids are found in complex active regions (ARs) often considered to contain a magnetic flux rope-like twisted structure (Titov & Démoulin 1999; Gibson et al. 2006a; Tripathi et al. 2009) in their corona. They appear to take the shape of a forward 'S' or an inverse-S, resembling a 'Z'. They make large observable features against the solar disk and are usually seen in soft and hard X-rays, as well as short extreme ultraviolet (EUV) wavelengths. Examples of sigmoids in soft X-rays taken using the X-Ray Telescope (XRT) onboard the Hinode spacecraft (Kosugi et al. 2007) are shown in Fig. 1. Forward-S and inverse-S sigmoids take their shape due to magnetic twist, and are characterized as right- and left-handed, respectively. Twisting of magnetic structures often leads to eruptive instabilities in the Sun (Cheung & Isobe 2014 & references therein) and can be measured by a quantity known as magnetic helicity, which quantifies the amount of twisting and writhing along magnetic field lines. Twist refers to the twisting of the field lines about the axis and writhe refers to the axis coiling around itself (Berger & Field 1984; Martin et al. 2012). Measuring helicity is useful for estimating when an eruption might occur and is a conserved quantity in environments of high magnetic Reynolds number (fraction of the characteristic speed of the plasma to the characteristic length over the magnetic diffusivity) and is thought to be "removed" from the Sun only bodily via helical ejecta, predominantly CMEs.

Sigmoids have been widely studied because of their complex morphology and their propensity for eruptive flaring (Canfield et al. 1999; Sterling 2000; Glover et al. 2000). Kusano (2005) suggested that they spontaneously form in regions where there is a shear inversion layer, that is where emerging fluxes of opposite shear can come in close contact with the preexisting ones, causing a growth in the resistive tearing mode instability. This essentially relaxes the energy buildup thus sustaining the structures forming the sigmoid without leading to an energy buildup and eruptions. Alternatively, Fan & Gibson (2003) suggested the possibility of sigmoidal structures emerging through the photosphere already twisted. They use this as a starting point in their models to study how the sigmoids evolve and produce eruptive events. Importantly, they show that the structure of the electric-current density or the current sheets in their sigmoid region are concentrated at the PIL. Although the current sheet is shown to be continuous and the sigmoid emerges as one flux tube in these simulations, high resolution observations show that they are made of multiple hot plasma features (Glover et al. 2000; Archontis et al. 2009) including short J- and S-shaped ones (Kliem et al. 2004; Pan et al. 2022).

Canfield et al. (2007) studied 107 ARs judged as containing sigmoids and obtained field directions via current-free (i.e. "potential") field extrapolations of the sigmoids at heights of 0.4-0.8 R$_\odot$ and that of the overlying arcade field at 2.5 R$_\odot$ to observe if they match in terms of their magnetic field directions. They found parallel and anti-parallel configurations between the two heights and, remarkably, that the ones with anti-parallel configurations flared almost twice as much as the ones with parallel configurations. They suggested two main models of reconnection that may be responsible for eruption in the anti-parallel and the parallel configurations. In the anti-parallel configurations, the eruption occurs due to the interactions between the overlying field and the sigmoid field via breakout reconnection(Antiochos 1998). Breakout reconnections can occur when there are multi-polar flux systems present in close vicinity forming a null point in the corona. Sigmoids usually form in regions of complex and multiple polarities. The flux systems interact at the null point leading to magnetic reconnection high in the corona opening up the overlying field lines and leading to an eruption. This is considered to be more efficient compared to tether-cutting reconnection (see an example in Aparna & Tripathi 2016). Canfield et al. (2007) suggest that tether-cutting reconnection is the possible mechanism in cases with parallel configurations between the overlying field and the sigmoid field.

Sigmoids are classified as transient and long-lasting (several days). Transient sigmoids are short lived, they often lose their S or Z shape post an eruption commonly developing a cusp like shape indicating magnetic reconnection leading to an eruption (Glover et al. 2000; Török & Kliem 2005; Gibson & Fan 2006; Vasantharaju et al. 2019). Long term Sigmoids may retain their S or Z shapes for several days and even after an eruption suggesting that the current sheet survives, potentially leading to more subsequent eruptions. Gibson et al. (2006a) suggested that sigmoids are large flux-rope structures and that the dipped parts of the sigmoidal flux ropes are seen in observations as 'S' or 'Z' shapes as those are the regions where tangential discontinuities can develop in the magnetic field giving rise to electric-current sheets heating the plasma to high enough temperatures that emit in X-ray wavelengths.

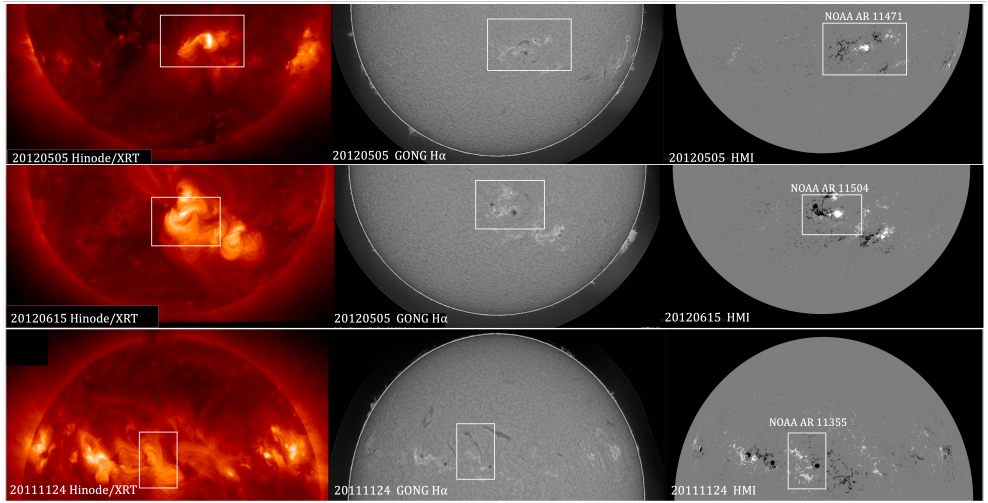

**Figure 1.** Selected sigmoids from the Hinode-XRT catalog with their filaments and the line of sight magnetogram.

### 1.2. *Filaments*

Solar filaments are dark structures seen commonly in absorption in Hα, He 10830Å, Ca II H & K and other chromospheric wavelengths against the solar disk, or in emission above the solar limb (in which case they are called prominences). They include relatively cold plasma of about 10,000 K thought of as suspended by magnetic dips at coronal heights. Filaments are often situated inside flux ropes, in other words, flux ropes often wind around the filaments (van Ballegooijen & Martens 1989). In white light limb images of CMEs in the corona, prominences are seen as bright features at the bottom of the "cavity" region (Gibson et al. 2006b), which is a dark region devoid of electrons required for scattering the photons. Above the cavity, there is the expanding loop of the CME corresponding to the overlying arcade loops seen on the disk with the cavity being the flux rope situated above the filament/ prominence. The bright prominence, dark cavity and overlying loops together form what is known as the three-part CME structure (Riley et al. 2008) often seen during CME eruptions. Filaments are made of threads of magnetic field lines along which plasma is observed to flow back and forth (Zirker et al. 1998; Litvinenko 2021). They always form above a photospheric PIL, also known as the "neutral line", precisely on the interface between the two polarities where the radial magnetic field component drops to zero.

Filaments are categorized into quiet region, intermediate and active region filaments depending on where they occur (Engvold 1998). Quiet region filaments are often found with distinguishing structures known as barbs that extend out of the central long structure called the spine. In AR filaments, the barb structures are usually absent. However, in all filament kinds, the fibrils that make up the adjacent spaces of the filament neutral line, follow along in a certain direction, that is suggestive of the "handedness" of the filament called as chirality, similar to the handedness of sigmoids represented by their shapes. Filaments are left (sinistral) and right (dextral) handed chiral structures when the tails of the chromospheric fibrils adjacent to the PIL point away from the PIL towards the left and right, respectively as seen from the positive polarity. Filament barbs if present, branch out away from the filament spine to the left and right, respectively as well. The chirality of filaments is governed by magnetic field, hence following along the fibrils by knowing the magnetic polarities from where they originate, can be used for obtaining the directionality of magnetic field along the filaments. This is important for space weather prediction because CMEs are commonly seen with prominences in their cavities and the directionality of the CME's axial magnetic field plays a decisive role in whether the CME will reconnect with the Earth's magnetic field to cause a geomagnetic storm or not. Using data sets from different years, previous works such as Burlaga et al. (1981); Bothmer & Schwenn (1998); Marubashi et al. (2015);

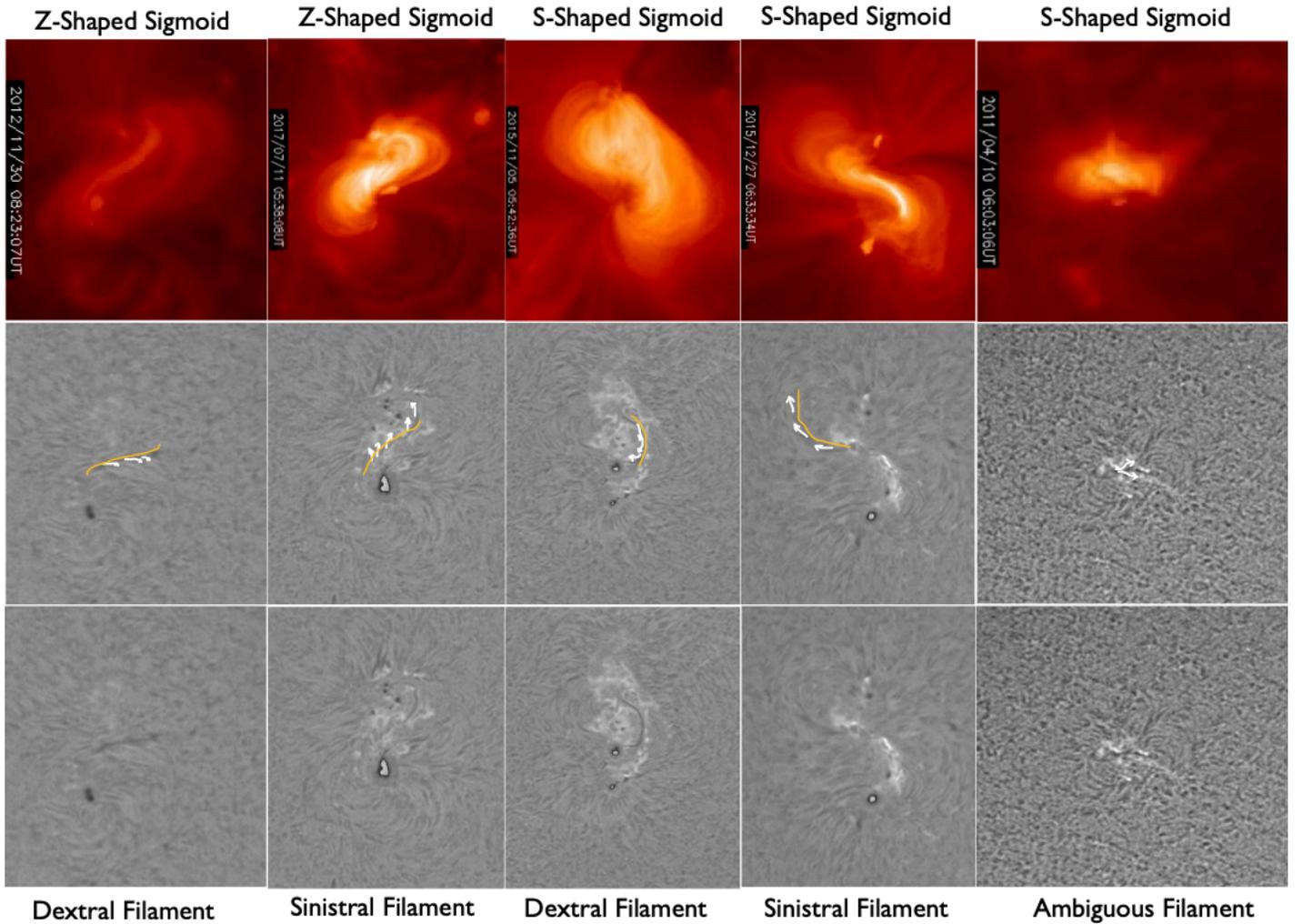

**Figure 2.** An example corresponding to each shape (S and inverse-S or Z) of a sigmoid with a filament dextral and sinistral filament are shown along with an example filament that is ambiguous. The top panel gives the X-ray sigmoids with the bottom panel showing the corresponding Hα filaments. The middle panel shows how the chirality of the filaments are determined with the white curves representing the direction in which the filament threads are turning away from the central spine shown in yellow. The Hα images are not scaled as per the resolution of XRT images.

Palmerio et al. (2018) and Nieves-Chinchilla et al. (2019) have indicated in essence that the axial field direction is mostly retained during their post eruption travel from the Sun to the Earth. Our previous works - Aparna & Martens (2020) and Mundra, Aparna, & Martens (2021) build up on these and use the largest data set of CME magnetic clouds (MC) at L1 Lagrangian point and filament counterpart of the CMEs on the Sun. Our data sets cover almost two solar cycles, namely from 1996 to 2017. We find that in about 75% of the cases, the axial field direction is retained between the Sun and the Earth indicating that observing the filament magnetic field direction can help predict whether an ensuing CME will or will not affect the Earth's magnetosphere.

In both our studies, we have used filament chirality to find the axial field directions. The first aim of the current work is to identify if a consistent relation exists between the handedness of the sigmoids and the corresponding filament chirality. This is essential to determine because, if sigmoids were to be as well-suited for prediction as filaments, the fact that they are large observable features would make it easier to use their shape for predicting the effect of an eruptive event on the Earth's magnetosphere, using automated algorithms. The sigmoid handedness in general complies with the hemispheric helicity rule. The filaments while erupting, may also experience a rotational motion in compliance with the hemispheric preference (Green et al. 2007). However, filaments in general have been observed to be less in compliance compared to sigmoids (Zhou et al. 2020).

### 1.3. *A note on the magnetic structure of the Sigmoids and Filaments*

The magnetic structure of filaments can be best understood by realizing that the magnetic field of and around the filament is dominated by a very strong horizontal component parallel to the polarity inversion line (for example, see Fig. 3 in Martin et al. 1994). Other observations confirming this, are the orientation of fibrils along the PIL (Foukal 1971), the shear of postflare loops after the filament erupts, and the shear angle of the overlying loops prior to the filament eruption.

The only observables that seem to contradict the dominant magnetic field direction are the filament barbs. They veer off from their filament at an angle that indicates that their parallel magnetic field is opposite to the dominant axial field of the filament. This apparent contradiction was resolved by Martin (1998a) and references therein, who observed that filament barbs are rooted in minority polarities, which implies that the magnetic field along their axes is opposite to what one would expect looking at a magnetogram, and hence consistent with that of their filament. We note that this magnetic structure of filaments dominated by the magnetic field parallel to their axes is reproduced by both main theoretical models for filaments, that of a flux-rope (van Ballegooijen & Martens 1989) and that of a sheared arcade (Antiochos et al. 1994), and hence it cannot be used to discriminate between the two. Both these models also imply that force-free filaments with positive (negative) chirality have a negative (positive) current helicity along their field lines as well as a negative (positive) total helicity for the filament body.

The magnetic structure of sigmoids is far less clear from their observations. Sometimes parts of the sigmoid look like flux tubes stretched out along their accompanying filament. Other parts of the same sigmoid can look like an arcade of loops with the arcade forming the body of the sigmoid. In this paper we find that at times the helicity of the filament and its sigmoid are of opposite signs, indicating two separate magnetic entities. Martin et al. (2012) suggested that the sigmoids exist in the cavity that is the space between the filament and the overlying arcade and has the same chirality as that of the overlying arcade. However, the causes of the 'S' and 'Z' shapes and the fact that these shapes are not seen in all active regions having a filament are intriguing and needs investigation.

There have been very few studies that have analysed filaments and sigmoids together. For example, Régnier & Amari (2004) used non-linear force-free field extrapolations and identified a) dips such as those in a Kuperus-Raadu type (Kuperus & Raadu 1974) magnetic field configuration at the locations of the Hα filament, b) a large twisted flux-tube structure that was at a height above the third structure, c) which corresponded to the sigmoid which formed around the filament with a turn under the filament. Similar to some of our filament-sigmoid pairs, the alpha and current density values in their sigmoid had opposite signs. Contrary to the observations of Pevtsov (2002) and the models of Gibson & Low (2000) who assume that both features correspond to the same structures, Régnier & Amari (2004) also suggest that they are two separate structures.

Our second aim is motivated by the results of the first part of this work which suggests that there is no consistent correlation between filament chirality and sigmoid handedness. In the second part, we explore how the handedness or helicity sign in the sigmoid structure is different or similar with that at the photosphere and filaments. We use the sigmoid handedness to visually determine the helicity sign in the corona, the chromospheric filament chirality to determine its helicity sign in the chromosphere, and we use calculations performed using the method of Georgoulis et al. (2012), to get the helicity sign of the sigmoid region in the photosphere. We find that the photospheric helicity sign matches sometimes with that of the filament and sometimes with that of the sigmoid. We give details of the results in the following sections.

The paper is organized as follows. We first discuss the data and analysis methods used for two aims of this work in section 2. Then we discuss the results from the analysis of the first aim, then we discuss the results of the analysis of the second aim in section 3. We conclude with some discussion in section 4.

## 2. DATA AND ANALYSIS

### 2.1. *Filaments and Sigmoids*

Out first aim is to identify X-ray sigmoids with their filament counterparts and compare their helicity and chirality signs. For a quick reference, we give a one to one mapping between chirality and helicity signs in Table 1 for these structures. Structurally, this essentially means that the topology of a flux rope for a filament with right oriented barbs is such that it wraps around the filament by going under near the barb locations such that it has a negative helicity for a dextral filament and vice versa (see Gibson & Low (1998) for example). This flux rope has a negative (positive) twist and hence negative (positive) helicity signs and the appendages or barbs can only be dextral (sinistral) with the overlying arcade of the filament skewed to the left (right), respectively.

| Filament Chirality | Helicity Sign | Sigmoid Shape | Overlying Arcade Skew |
|---|---|---|---|
| Dextral | − (Left) | Z | Left |
| Sinistral | + (Right) | S | Right |

**Table 1.** Quick lookup table to show the sign of filament chirality, corresponding helicity sign, and shape of the arcade overlying the polarity inversion line (see Martin (1998b)).

To compare the two signs, we have built a dataset consisting of Hα filaments and co-spatial soft-Xray sigmoids from 2007 to 2017 containing their chiralities and helicities. A similar analysis was conducted by Martens et al. (2014) for sigmoids and filaments between Oct 2010 and Mar 2011. Our list of the sigmoids is obtained from Savcheva et al. (2014)[1] and their continued analysis of sigmoid detection and measurements from 2013-2017[2]. Their data is available until 2017. To obtain the chirality of the filaments, we mainly use GONG Hα images and also Kanzelhöhe Solar Observatory (KSO, Pötzi et al. 2015) Hα images, in a few cases where GONG data were unavailable. Using these images, we visually determine the chirality of filaments and the shape of the corresponding sigmoids using daily images from the Hinode XRT sigmoid calendar[3] obtained from the X-Ray Telescope onboard the Hinode spacecraft (Golub et al. 2007). The filaments lie within the region of the sigmoids, above photospheric PILs. We use the images of Hinode XRT sigmoids for the days present in Savcheva et al.'s catalog and the chiralities of the filaments are analysed for the same days. We obtain the filament chirality using the methods described in Martin (1998b), mainly using filament barbs and fibrils adjacent to the filaments. Detailed description of the method can also be found in Aparna & Martens (2020). If filament chiralities are not clear on the days of sigmoid detection, we obtain the chiralities from observing the same region but on different days by making an educated assumption that the field configurations do not reverse in such small time-scales and that the filament in the studied active regions will maintain their chirality over different days. To our knowledge, filament chiralities do not change over the lifetime of the filament.

In a noteworthy study, Ouyang et al. (2017) have performed an extensive statistical analysis of 571 filaments and suggest that Martin's chirality method only applies when the filament region has an inverse magnetic polarity configuration (Kuperus & Raadu 1974). Ouyang et al. propose that Martin's rule applied to filament regions in a normal polarity configuration will give a chirality opposite to that in reality. We do not necessarily agree with these results. Instead, we think that both methods of deriving chirality should yield the same results (Martens & Aparna, in prep). Our analysis has been performed based on Martin's methods and the results of comparing filament chirality with the sigmoid handedness for 85 sigmoids between 2007 and 2017 are given in Table 2.

## 2.2. *EnHel*

### 2.2.1. *The α Parameter*

Our second aim is to quantitatively assess the helicity sign using photospheric magnetic field as the lower boundary and then compare it with that of the filaments that form in narrow filament channels above polarity inversion lines, and with the shapes of sigmoids, from visual analysis. We compare helicity signs of the observed sigmoid region using XRT data, chirality of filaments in the region above the PIL derived from Hα filament observations and the photospheric helicity sign which is obtained from calculations over the region of the sigmoid AR using photospheric vector-magnetograms. For the photospheric calculations we use SHARP (Space-Weather HMI Active Region Patch; Bobra et al. 2014) vector magnetograms from the Helioseismic Magnetic Imager (HMI; Scherrer et al. 2012) onboard the Solar Dynamics Observatory (SDO; Pesnell et al. 2012). We obtain the SHARP data formatted according to the Lambert Cylindrical Equal Area (CEA) projections of the magnetic field vectors from the JSOC database[4]. Using vector magnetograms as inputs we implement the non-linear force free (NLFF) method of Georgoulis et al. (2012), hereafter referred to as EnHel, to get the dominant sign and value of the force-free parameter α, for the sigmoid active regions in the photosphere.

---

[1] aia.cfa.harvard.edu/sigmoid/

[2] aia.cfa.harvard.edu/sigmoid23-24/sigmoid_webpage/

[3] http://solar.physics.montana.edu/HINODE/XRT/SCIA/latest_month.html

[4] http://jsoc.stanford.edu/

EnHel uses all three photospheric magnetic field components as input in Cartesian or polar coordinate systems and calculates instantaneous magnetic energy and helicity budgets corresponding to the studied vector magnetogram. More on energy and helicity computations is given in subsection 2.3. The EnHel method of Georgoulis et al. (2012) is a connectivity based model that can be used for both linear (Georgoulis & LaBonte 2007) and non-linear force-free calculations to get the lower limit of the instantaneous energy and helicities of an active region using measured photospheric vector magnetic field. They use a flux tessellation scheme based on Barnes et al. (2005) to build a connectivity matrix between the positive and negative polarity patches. To form the connectivity matrix, a minimum flux threshold, minimum number of pixels (i.e. area) in each patch area and a threshold vertical magnetic field strength are assigned. This is done so that it includes only relatively strong magnetic fields that cannot belong to nearby quiet-Sun regions. The selection of flux patches is then subjected to a simulated annealing scheme aiming to minimize the distance between the footprints of magnetic flux tubes (hence emphasizing magnetic PILs) and magnetic flux imbalance. The fluxes are considered to be balanced if the absolute values of the positive fluxes and negative fluxes in a patch of active region are almost the same. The EnHel code requires the fluxes to be balanced to satisfy the principle of field conservation and avoiding large uncertainties in the connectivities. The photospheric-coronal magnetic continuum is then translated into a discrete set of flux tubes, that are supposed to be slender, connecting positive- and negative-polarity flux patches. The force-free parameter $\alpha$ and its error is calculated for each flux tube connection. We obtain the flux-weighted average $\alpha$ and its error $\delta\alpha$ to represent the prevailing helicity sign of the active region and compare it with the observed helicity signs. The flux-weighted $\alpha$ and error in $\alpha$ are obtained as follows.

$$\alpha = \frac{\sum_k \alpha_k \phi_k}{\sum_k \phi_k} \tag{1}$$

$$\delta\alpha = \frac{1}{N}\sqrt{\sum_{k=1}^{N}\Delta\alpha_k^2} \tag{2}$$

where, $\alpha_k$ and $\phi_k$ are the force-free parameters and the flux of each slender flux tube, respectively, connecting the patches in a magnetogram. We use the NLFF version of the EnHel code with an operational threshold of $5 \times 10^{19}$ Mx for the minimum flux for a patch, 30 magnetogram pixels for minimum patch size (i.e., ~0.16 $Mm^2$), and a 50 G lower-threshold for the vertical magnetic field per pixel.

### 2.2.2. *Flaring Propensity*

For the sigmoids in our dataset, we consider the results for only those active regions with a flux balance better than 20%. The list of our sigmoids with the properties derived visually as well as from EnHel are published at the Harvard Dataverse (Aparna 2023). To save space, we present a subset of these properties in the current manuscript in Table 4. It starts with the index and we give the date of sigmoid detection obtained from Savcheva's catalog in the second column, the third column gives the shape of the sigmoid, Z or S, followed by the helicity sign, + or -, respectively. The fifth column gives the chirality of the filament in the sigmoid region, L and R for left- and right-handed chiralities. The helicity signs corresponding to these chiralities are given in the sixth column. S, N and A in the fifth column refers to small, no (observed chirality) and ambiguous (chirality) filaments, respectively. The seventh column gives the flux-weighted $\alpha$ in units of $Mm^{-1}$ obtained from the EnHel code, followed by the uncertainty in $\delta\alpha$ measurement in the eighth column. Twist of a flux tube is a self-property showing rotation about its own axis (self-helicity) and mutual helicity is due to the interaction of adjacent, self-helical flux tubes. To obtain the number of turns in the sigmoids, we use the length of the sigmoid from Savcheva's catalog and multiply by the calculated alpha value, in the ninth column. The tenth column gives the flux balance percentage. The eleventh column shows flare-index calculated based on the the number of flares from the respective sigmoid region. The flare records were obtained from the Solar Monitor[5] and the occurrences of CMEs were identified visually by overlaying AIA, HMI and LASCO C2 and C3 images using JHelioviewer (Muller et al. 2009) as well as using the CDAW halo-CME catalog[6] (Gopalswamy et al. 2009). Very few (~8) halo-CMEs have been recorded from these sigmoids. The last column in Table 4 gives the coordinates of the location of the feature on the Sun in the heliographic system.

---

[5] solarmonitor.org

[6] cdaw.gsfc.nasa.gov/CME_list/halo/

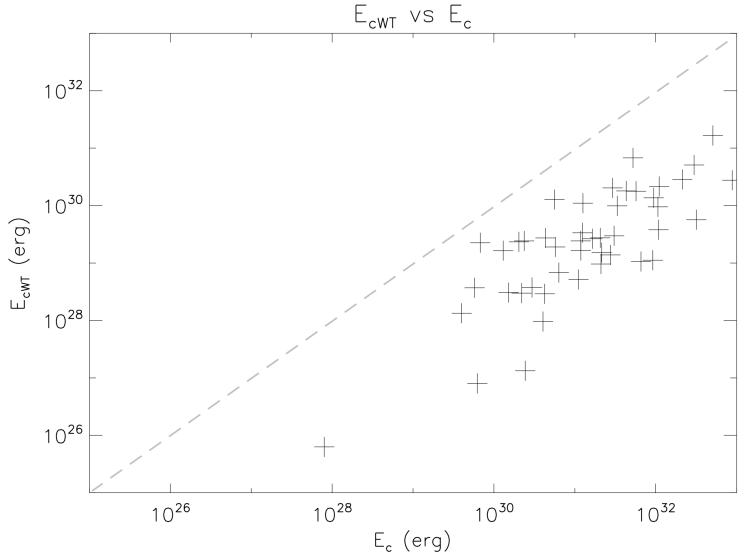

**Figure 3.** Free-energy correlated to the Woltjer-Taylor minimum energy for the set of 46 sigmoids. The gray dashed line signifies that all NLFF free-energies are higher than the Woltjer-Taylor energies.

We have 46 sigmoids in our catalog that also have an Hα filament in them, however, four of these had filaments with non-detectable chirality. But we have run EnHel on all 46 of them. It was not feasible to run all the sigmoids for multiple days mainly because of computational time. As a sanity check for the calculations performed, we show a comparison of the calculated free energy with respect to the respective Woltjer-Taylor minimum energy (namely, the linear force-free magnetic free energy for a given nonlinear force-free magnetic helicity as shown in equation 6) in Fig 3 for the 46 sigmoids.

We perform EnHel calculations for each sigmoid for a 60 minute interval (from 12:00 UT), at full HMI-SHARP cadence of 12 minutes, that provides us with six energy/ helicity values from the start of an hour to the start of the next. From these values, we choose the maximum helicity and its corresponding free-energy to conduct the analysis. In order to understand the relation between the sigmoids and their flaring propensity, we analyse the number of flares and CMEs associated with the 46 active regions by obtaining their flare indices (Abramenko 2005) using the following formula.

$$FI = \frac{100\sum_i X_i + 10\sum_j M_j + 1\sum_k C_k}{D} \tag{3}$$

Here, FI is the flare index, $X_i$ represents the scale of the GOES X-class flares, with index i running through their number of flares, while $M_j$ and $C_k$ correspond to all cases of NOAA M-class flares and C-class flares, respectively. D is the number of days over which the flares are counted. The values of {1,10,100} are the weights given to {C,M,X} flares such that the FI has units of $10^{-6}Wm^{-2}Day^{-1}$. The FI is calculated from the day after the sigmoid is detected until the AR remains on the solar disk (i.e., Earthward solar hemisphere). The flare indices calculated accordingly are given in the last column of Table 4.

### 2.3. *Analysis of a Long-term Sigmoid*

Some sigmoids are known to last for several days (Mulay et al. 2021) while some others are transient (Archontis et al. 2009) lasting from a few minutes to a few hours just before eruptions occur (Gibson et al. 2006a). We have chosen one long-lasting sigmoid from December 2015 that lived on the disk for almost the entire transit-time of the source AR on the solar disk, and performed EnHel calculations. We analyse how the sigmoid configuration may be related to the eruptive events it produces and how it might be related to the force-free values and the flux emergence. Our intention was to perform such calculations for multiple long-lasting sigmoids. However, due to a lack of such long-lasting sigmoids, we defer this analysis to a future time.

The NLFFF calculations from EnHel not only provide the sense and values of the twist in the active region structures which we use for comparing the helicity signs, they also give the potential field, right, left and the total magnetic helicity values, Woltjer-Taylor minimum energy and the free-energy stored in the system, unsigned and total fluxes in the system, etc. We study the evolution of these quantities for one long-term sigmoid of Dec 2015. The formulae that EnHel uses for calculating these quantities are as follows.

For each set of vector magnetic field input, EnHel outputs potential energy ($E_p$), total magnetic energy - $E_t = E_p + E_c$ where $E_c$ is the free magnetic energy given by,

$$E_c = Ad^2 \sum_{l=1}^{N} \alpha_l^2 \Phi_l^{2\lambda} + \frac{1}{8\pi} \sum_{l=1}^{N} \sum_{m=1, l \neq m}^{N} \alpha_l \mathcal{L}_{lm}^{arch} \Phi_l \Phi_m \qquad (4)$$

and, the total magnetic helicity $H_m$ is given by,

$$H_m = 8\pi A d^2 \sum_{l=1}^{N} \alpha_l \Phi_l^{2\lambda} + \sum_{l=1}^{N} \sum_{m=1, l \neq m}^{N} \mathcal{L}_{lm}^{arch} \Phi_l \Phi_m \qquad (5)$$

The first and second term in the energy and helicity equations above correspond to the self and mutual helicity terms due to individual flux tubes and their interaction with the different surrounding flux tubes, respectively. The self terms are obtained from the potential field calculations of LaBonte et al. (2007) which serve as the main basis for EnHel (Georgoulis et al. 2012) and are extended to suit the NLFFF type calculations. The mutual terms have been derived using the study of Demoulin et al. (2006). In this study, we mainly use the total, left and right helicity values to study the evolution. The mutual helicity and energy term generally dominate the self terms as already been shown by Georgoulis et al. (2012), hence we do not study them separately. The left and right helicities are the total helicities with a negative and a positive sign, respectively.

The Woltjer-Taylor minimum energy which acts as the lower limit for the free-energy under linear force-free conditions, is used for sanity-checking the EnHel calculations. It is calculated as follows,

$$E_{c_{WT}} = \frac{H_m^2}{(8\pi d^2)\mathcal{F}_{lin}E_p} \qquad (6)$$

where, $d$ is the pixel scale and $\mathcal{F}_{lin}$ is a linearized scale factor calculated in Fourier space, from 56,686 MDI (Michelson Doppler Imager; Scherrer et al. 1995) magnetograms, between $\mathcal{F}_{lin}E_p$ and $\Phi^2$ where $\Phi$ is the unsigned magnetic flux. We show the results of the analysis from the three parts explained above in the following section.

## 3. RESULTS

The two main results of this work are described in the following subsections. Sections 3.1 and 3.2 provide the results of the statistical analysis from sections 2.1 and 2.2, and section 3.3 focuses on a select case of a long-term sigmoidal active region NOAA AR 12473.

### 3.1. Relation between Filaments and Sigmoids

From the first part of the analysis, we recall that the chirality of filaments and the handedness of the sigmoids do not always have the same sign; a dextral filament can be present in an 'S' shaped sigmoid and a sinistral filament may be present in a 'Z' shaped sigmoid. In other words, large scale sigmoids seen in coronal (X-ray and EUV) wavelengths, consist of filaments that may be of helicity opposite to that of the sigmoid. Figure 2 shows examples of 'S' and 'Z' shaped sigmoids each containing filaments of dextral and sinistral chirality. An example of an ambiguous filament in an 'S' shaped sigmoid is also shown. The results of the comparison of filament chirality and the sigmoid helicities between 2007 and 2017 are tabulated in Table 2 and are described below.

- There are 103 recorded sigmoids in total from 2007 to 2017. Of these, 52 are forward sigmoids and 51 are inverse sigmoids. 41 of the forward and 44 of the inverse sigmoids have Hα filaments in them. 16 of the 44 inverse sigmoids have dextral filaments in them and about the same amount (∼14) have sinistral filaments in them. 21 of the 41 forward sigmoids have sinistral filaments and eight of them have dextral filaments. The remaining amounts in each category have ambiguous filament chiralities. The sizable departure in the helicity sign of the filaments in inverse sigmoids is noteworthy.

- While there is slight preference for dextral (sinistral) filaments to inverse (forward) sigmoids, from a space weather prediction perspective, we find that a filament of either chirality could reside in a sigmoid of either shape. We note that higher resolution filament images from Goode Solar Telescope and the upcoming Daniel K. Inouye Solar Telescope (DKIST) will potentially improve the statistics of the result.

| Sigmoids → | Forward (S) | Inverse (Z) |
|---|---|---|
| Filaments ↓ | (41) | (44) |
| Dextral | 8 (20%) | 16 (36%) |
| Sinistral | 21 (51%) | 14 (32%) |
| Ambiguous | 12 (29%) | 14 (32%) |

**Table 2.** Distribution of filaments and sigmoids according to chirality and shape respectively for data between 2007 and 2017. We remind the reader that agreement implies dextral (sinistral) filaments associated with inverse (forward) sigmoids. Ambiguous refers to the filament cases that were unable to be determined for left and right chiralities.

- 75% of our 85 sigmoids comply with the hemispheric preference while only 54% of the filaments follow the hemispheric preference.

### 3.2. *EnHel Results*

From the second part of the analysis involving photospheric vector-magnetograms, we use 46 of the 55 sigmoids between 2010 and 2017 selected for a run with EnHel. Nine of the 55 had a flux balance ratio much higher than 20% and are omitted. The distribution is summarized in the following points, and a visual representation is given in Fig 4. The results are categorized based on the helicity signs, hemispheric preference and flaring propensity and are elaborated under the respective category below.

1. Helicity Signs —

   (i) Of the 46 sigmoids, 22 cases show a match in the helicity signs between two of the three $\alpha$ signs pertaining to photospheric calculations, filaments and sigmoids. The first cells of the rows corresponding to these cases are highlighted in Teal in Table 4.

      - Of the 22, nine cases have helicity signs that match between the sigmoids and the average alpha (calculations).
      - 13 of them have helicity signs that match between the filaments and the calculations.

   (ii) Of the 46 sigmoids, 16 cases match in the helicity sign between all three - filaments, sigmoids and EnHel calculations. The first column of the rows corresponding to these cases is highlighted in Purple in Table 4.

   (iii) Four of the 46 cases have the same observed helicity sign in the sigmoid and the filament, however the EnHel calculations give the opposite sign. We exclude these four sigmoids while comparing the flare outputs from the full-match and partial-match categories. These rows are shown with green color in Table 4.

   (iv) Four of the 46 sigmoids have filaments with unidentifiable chirality and are colorized with gray.

2. Hemispheric Preference —

   (i) For the cases where there is a match in helicity signs between all the three features, the hemispheric helicity preference holds in 14 of the 16 cases (88%).

   (ii) For the cases where there is a match between any one of the helicity signs (of filament or sigmoid) and the EnHel calculations (from photospheric vector fields), the hemispheric helicity rule holds only in 14 out of 22 cases (64%).

   The hemispheric preferences for the sigmoids in the two cases are tabulated in Table 3. The names - Purple and Teal Sigmoids are based on the color coding for the respective sigmoid rows as shown in Table 4, those that match with all three features (purple) and those that match the calculations with one of the features (teal). The data is also available in the online data-sheet in Harvard Dataverse (Aparna 2023). The results described here are also as depicted in Fig. 4.

3. Flaring Propensity —

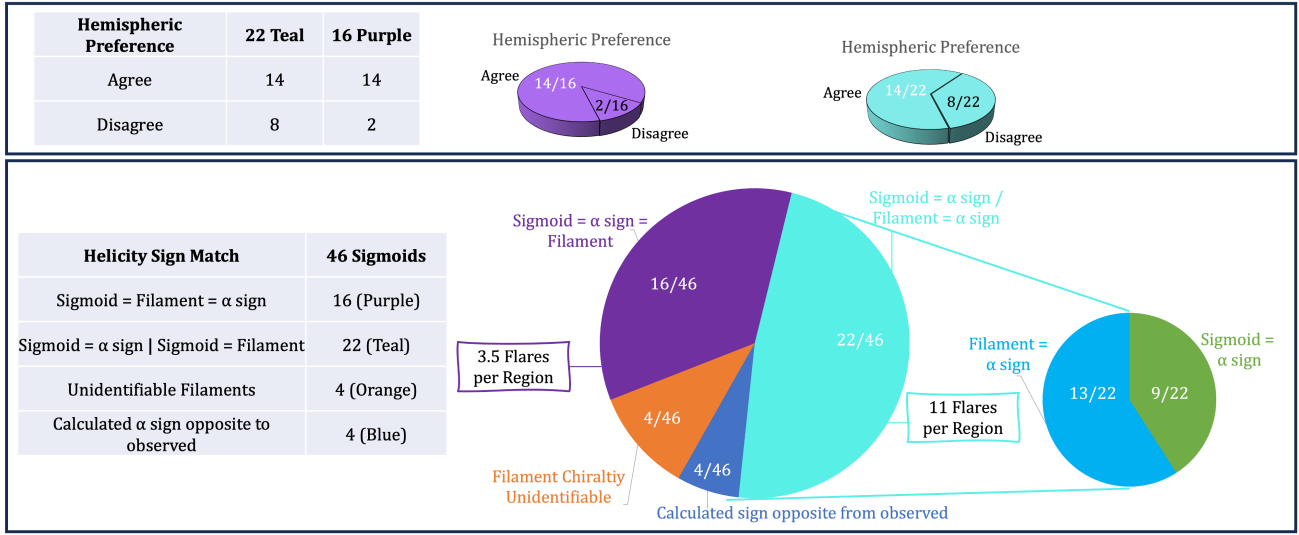

**Figure 4.** Results comparing the helicity signs from EnHel calculations (using photospheric vector-magnetograms), and the observations of filament chiralities and sigmoid shapes. Larger slices in the biggest pie represents sigmoids with EnHel matching observed helicity sign with the filaments or the sigmoids (in Teal), and sigmoids matching all three signs (in the purple slice). The smaller pie on the below right connecting the teal sigmoids represent the number of sigmoids in each of its categories. The smaller pies at the top represent the hemispheric helicity preference for the purple and teal sigmoids. The tables on the left of the pie-charts summarise these numbers as well. A more detailed distribution for the hemispheric preferences can be seen in Table 3.

| Purple Sigmoids ↓ | N. Hem. | S. Hem. |
|---|---|---|
| Forward (S) | 0 | 7 |
| Inverse (Z) | 7 | 2 |

| Teal Sigmoids ↓ | N. Hem. | S. Hem. |
|---|---|---|
| Forward (S) | 5 | 6 |
| Inverse (Z) | 8 | 3 |

**Table 3.** Tables show the hemispheric preference for the sigmoids where the helicity sign is a match in all three features (called purple sigmoids, in the left table) and those with that match partially (right table, called teal sigmoids). N. and S. refer to Northern and Southern hemispheres, respectively. According to the hemispheric preference, inverse (forward) sigmoids are normally observed in northern (southern) hemispheres. Please refer to Table 4 for the Purple and Teal sigmoids.

(i) From the 16 cases with matches between all three features, we have listed 57 flares, so the average flare productivity is 3.5 flares per region. For the 22 cases with partial agreement, we have listed 250 flares, so the average productivity is 11.4 flares per region. There are considerably more number of flares when the helicity signs of sigmoids and their filament counterparts are opposite.

(ii) In general, in the cases with all matches, three out of 16 (∼19%) regions produced one CME each. In the cases where there is a partial match, six out of the 22 cases (∼27%) produced one CME and the rest did not have CMEs that were cataloged. In these cases in this dataset, the probability of CME production in the partial-match case versus full-match is about 1.5 times and that of flare production is about 3.3 times.

Following Fig. 2 from Tziotziou et al. (2012) we present the energy-helicity (EH) diagram for the sigmoids for our dataset in Fig. 5. It plots pairs of maximum magnetic helicity and corresponding free energy values in the time series of each sigmoid in logarithmic scale. Our sigmoids agree with the active region EH diagram from Tziotziou et al. in that the helicity and free energy are related by a power-law. The colorbar in Fig. 5 indicates the flare index. There are around 28 sigmoids that produced no flares during the time considered for calculating the flare-index, shown with dark-blue dots. And similar to theirs, our EH diagram shows a segregation between flaring and non-flaring sigmoids with most of the flare-producing sigmoid regions situated in the top right as opposed to non-flaring ones towards the left. We group our EH diagram based on flare indices of the sigmoid regions with the demarcating line in pink separating the points with low flare indices and those with high flare indices; the free-energy and total helicity cutoff for eruptive and non-eruptive ARs are $4 \times 10^{31}$ erg and $2 \times 10^{42}$ $Mx^2$, respectively. The sigmoids in Fig. 5 somewhat agree with these cutoffs as well, although the region with the highest flare-index falls below these limits. We obtain a threshold slightly less than the ones mentioned above, with a helicity of $1.1 \times 10^{41} Mx^2$ and an energy of $2.2 \times 10^{31}$ erg. It is important to note here that the helicity and energy values are only representative as they are calculated at a certain time and only the flares produced after that are considered. A more accurate helicity-energy diagram would include values at several times during the sigmoid's appearance, this will be attempted at a future time.

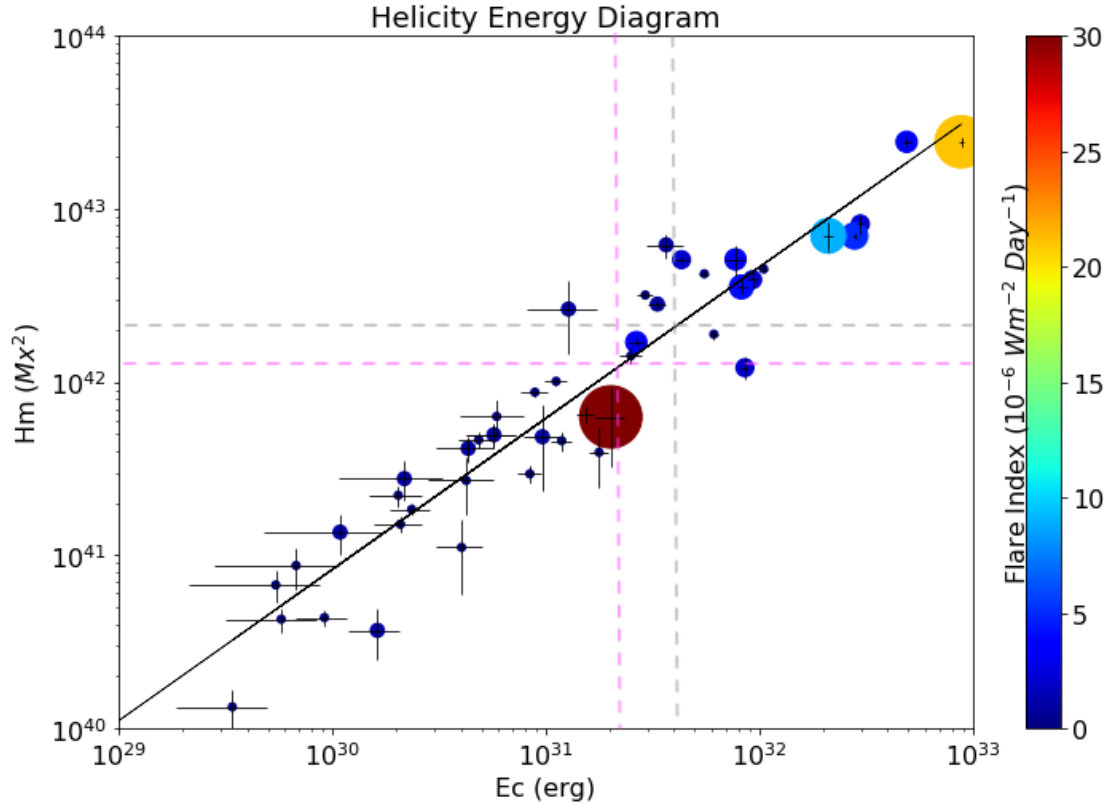

**Figure 5.** Energy-Helicity Diagram for 46 sigmoids of this dataset. Colorbar depicts the flare indices calculated using equation 3. Increasing size of the circles corresponding to increasing values of the flare indices. The black line indicates a linear fit with coefficients [0.87, 14.72] and the pink lines depict the cutoff for sigmoids with low and high flare indices. The gray lines are those obtained by Tziotziou et al. (2012)

The flaring sigmoids are associated with higher energies compared to the non-flaring sigmoids. However, there are also flaring sigmoids with free energy less than the cutoff energy stated above. This may indicate that these regions have a higher helicity balance i.e. sizable but nearly equal and opposite values of left- and right- handed helicities. We see that most of the sigmoids that produce more than 20 flares (with high flare indices) during their lifetimes fall above the cutoffs for $E_c$ and $H_m$. All the sigmoids producing zero flares fall below either of the cutoffs.

In Fig. 6, we show the helicity and energy values against the calculated absolute $\alpha$ values according to the flare index. The $\alpha$ values range between $\pm 2 \times 10^{-10} Mm^{-1}$. The distributions are fairly uniform between the sigmoids that give out different numbers of flares. Hence, the flux-weighted alpha values from EnHel do not necessarily indicate any patterns of flaring behaviour.

### 3.3. Sigmoid of December 2015 - NOAA AR12473

In order to better understand how eruptive events occur with changes in helicity and energy in the sigmoids, we perform hourly EnHel calculations for one sigmoid from December 2015 that lasted on the disk from the eastern limb to almost until the western limb.

NOAA AR# 12473 is seen on the east limb first on 22 Dec 2015. It attains sigmoidal form as early as on December 23, 2015 and survives almost until it reaches the western limb. It produces as many as 61 flares and six CMEs during this time. We perform EnHel calculations on the sigmoid region from Dec 23, 2015 until Jan 01, 2016 to understand how the changes in the energy and helicity relate to the eruptions. The active region in its entirety is not visible outside of these dates. We rely on the CEA data of the Bx, By and Bz fields from SHARP vector magnetograms for all these dates to perform the calculations. The calculations are performed every hour, of each day except on Dec 27 when the data for the entire day are not available. We show the variations of the different parameters pertaining to the Dec 2015 sigmoid over the time of its existence on the disk of the Sun in Fig. 7.

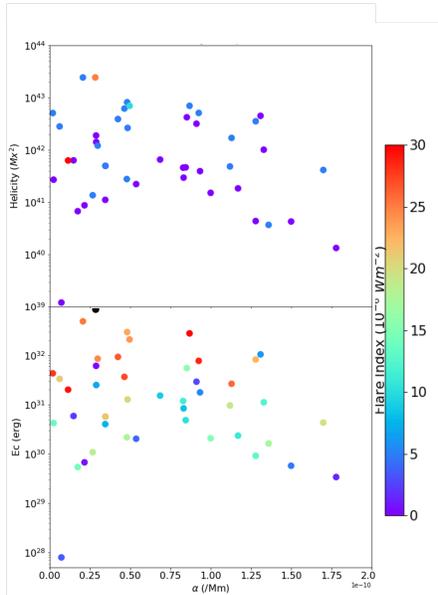

**Figure 6.** Distribution of helicity and free energy values for the 46 sigmoids against the weighted average of the force-free parameter.

The top panel of Fig. 7 shows the evolution of the magnetic helicity (left-, right-handed and total), followed by that of the $\alpha$ parameter, flux and the free energy. Each of the data series shown in open diamonds is smoothed and represented with solid lines. The smoothing is done using the IDL program TS_SMOOTH. The panel (third from the top) containing the flux evolution shows unsigned (pink), positive (red), negative (blue) and connected (black) fluxes. The bottom most panel gives the evolution of total free and Woltjer-Taylor minimum energies. The Woltjer-Taylor minimum energy is the minimum energy that a magnetic field configuration can hold for a certain helicity (Kusano et al. 1995). This is shown in green, and it has a very small variance compared to the total free energy plotted in purple with the latter's corresponding smoothing shown in pink. The changes in the free energy follows the trend similar to that seen in the changes in the helicity.

We choose the panel showing the $\alpha$ values as a guide to follow the evolution. The panel depicting the evolution of $\alpha$ shows a sizable valley around the middle of the plot of the time interval. $\alpha$ changes sign two times once from the positive to the negative around Dec 26th after which it becomes increasingly negative until around Dec 27th, and then from negative to the positive around Dec 30th after which it becomes increasingly positive. Fourteen flares occur prior to the start of the decline of the $\alpha$ values. Eight flares occur during the descent or the absolute increase in the negative values of $\alpha$. There is a data gap during Dec 27 after which there is a rise in the positive values of $\alpha$ and it continues to rise until Jan 01, 2016. Seventeen flares occur during the rise phase of $\alpha$ post Dec 27. The parameters in the remainder of the panels follow the $\alpha$ curve somewhat similarly. Details pertaining to the other parameters are as follows.

The top panel shows the evolution of total, right and left helicities. The total and right helicities show an initial increase in positive helicity until Dec 25 followed by a reduction until Dec 27 around which time, the total helicity changes sign similar to that seen in the $\alpha$ evolution. The magnitude of $H_m$ increases with a change of sign again seen around Dec 30th after which it increasingly becomes positive. The right helicity can be seen increasing around this time interval as well. The unsigned flux (in pink in the third panel) shows a gradual decrease between Dec 23rd and Dec 27th and a rise after Dec 29th. The negative (in blue) and positive (red) fluxes follow each other well, however, there is a higher rate of reduction in the negative flux compared to the positive flux between Dec 24th and Dec 26th. The flux emergence continues to increase from Dec 28 until the end of the plot-time on Jan 01. The flux decrease in the initial interval, before the plateau, appears to be taking away helicity that agrees with the hemispheric helicity rule. The plateau indicates helicity opposite to that of the hemispheric rule and then the active region returns to the helicity sign expected for its hemisphere.

The increase in the number of flares (8) between Dec 25th and Dec 27th perhaps can be related to the increase in the amplitude of the negative alpha parameter. We observe a decrease in the right handed helicity (top panel) while the left handed helicity values remain quite the same during this time. There is also some reduction in the negative fluxes during this period.

Similar activity (9+ flares) as above occurs between Dec 28 and Dec 30, when the $\alpha$ parameter gains an increasingly positive value along with an increasing right-handed helicity while the left-handed helicity remains as is. During this period, there is also continued flux emergence which is possibly adding both right- and left-handed helical structures to the system with right being more than the left and some of which might be cancelling with that of left. This however needs further investigation by observing the features at smaller scales. We conjecture that the initial reduction in the fluxes, $\alpha$ and helicity, between Dec 25 and Dec 27, perhaps indicate helicity annihilation (Kusano et al. 2004; Georgoulis et al. 2019) possibly triggering the flares during this period. The increase in the eruptive activity in the rise phase between Dec 28 and Jan 01, is possibly due to new fluxes adding more helicity and further annihilation. However, we do not have evidence for this phenomenon currently and further studies will be conducted to understand this.

## 4. DISCUSSION AND CONCLUSIONS

### 4.1. *Filaments & Sigmoids Comparison*

From the first part of the analysis, we find that the chirality of the filaments and the handedness of the sigmoids do not always have the same sign; a dextral filament can be present in an 'S' shaped sigmoid and a sinistral filament may be present in a 'Z' shaped sigmoid. In other words, large scale sigmoids seen in coronal X-ray and EUV wavelengths, consist of comparatively smaller scale structures such as filaments of opposite helicity, which frequently determine the direction of Bz of an erupting CME (Aparna & Martens 2020; Mundra et al. 2021). It is also possible that the filament and sigmoid structures are sufficiently disconnected from each other magnetically. Both the possibilities imply that the shape of these large scale structures (sigmoids) cannot be used for determining the magnetic field direction of ICMEs to predict space-weather effects on the Earth's magnetosphere. However, this result has implications to monitoring these regions for space-weather effects due to their flaring propensity as explained in the next subsection. This result also has implications for modelling of sigmoids consisting of emerging flux tubes that host prominences. Sigmoids are often depicted as twisted flux tubes and if prominences are included, they often carry the same sign of the twist (Gibson 2018 and references therein). While structures with similar twist signs are expected, our results suggest the need to explore models of prominences with opposite twists for insights into magnetic structure stability and eruptive activities.

We conjecture that the opposite twists in such examples are a result of two separate structures in the same volume. Although the filament is in the same magnetic field environment as the large scale sigmoid structure, the area near the filament is evolving much faster than the overall region which is mostly stable. But at the location of the filament, the PIL is a flux emergence region which perhaps brings about helicity fluxes that may be opposite to the dominant helicity of the overall sigmoid structure. It is out of scope of the current work to investigate this aspect which we plan to do in the future by contouring the PIL region and estimating helicity for that region separately from the overall region. Given that the sigmoids form in complex multi-polar active regions, and the filaments are a smaller subset of this complexity, it is possible that filaments are rooted in different flux patches in multi-polar structures which have different helicity signs than sigmoids.

### 4.2. *Filaments, Sigmoids & EnHel*

In the second part of our study, we perform energy-helicity type surface calculations using the code developed by Georgoulis et al. (2012) and compare the calculated helicity signs of the photospheric sigmoid AR with that of the observed helicity signs from H$\alpha$ filaments and X-ray sigmoids. The results from this part are summarised below.

- We have two cases, one where the helicity sign is the same between the photospheric calculations from EnHel, and those observed from sigmoids and filaments (42% of the 38). And the second where the photospheric calculations match with either the filaments or the sigmoids (58% of the 38). We call these full-match and partial-match cases respectively.

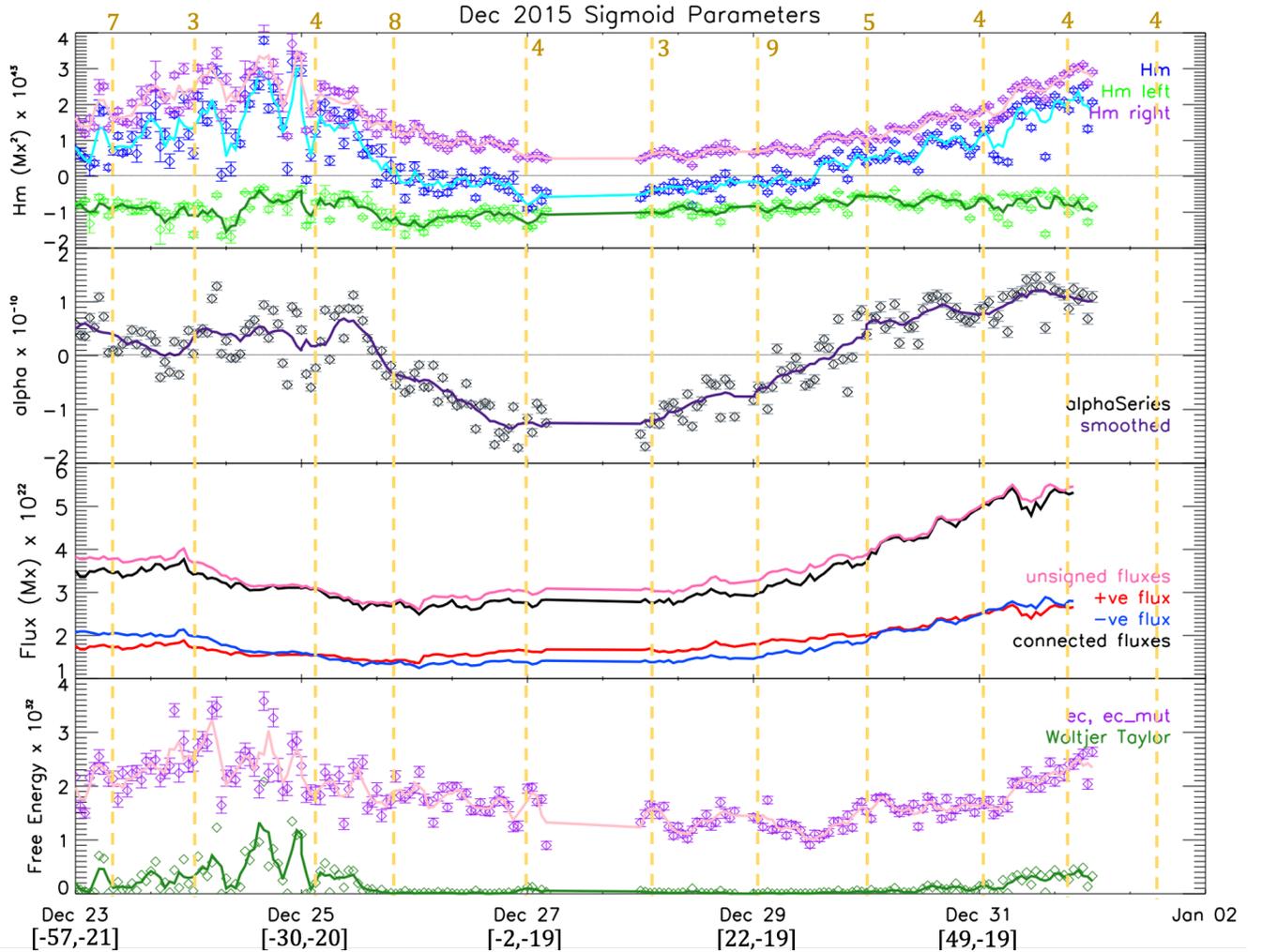

**Figure 7.** From the top: evolution of magnetic helicity, force-free parameter, flux and free energy for AR12473 between Dec 23, 2015 and Jan 01, 2016. Vertical dashed orange lines with numbers on the top axis indicate the number of flares that occurred on that day. Legend for the colors are shown within each panel. The continuous curves in the top, second and the last panels represent the smoothed data points. Heliographic position of the centroid of the sigmoid region over the time of disk passage is given in square brackets below the dates along the abscissa.

- Among the second kind in the point above, 13 of the 22 have the same helicity signs between filaments and sigmoids and nine of them have the same sign between the photospheric EnHel calculations and the sigmoids.

- The cases with a partial match produce about 3.5 times more flares than the cases with full match. The number of CMEs produced by the cases of partial match is about 1.5 times that produced by the cases of full-match.

### 4.2.1. *Hemispheric Preference*

The filaments in this study have a lower rate of agreement (54%) compared to their sigmoid counterparts (75%) for the hemispheric preference for all of the studied data between 2007 and 2017. This result compares well with previous studies where, by studying rotational direction of erupting filaments from sigmoidal regions it has been shown that both filaments and their sigmoids exhibit a forward S shape in the southern hemisphere and a reverse S shape in the northern hemisphere, indicating weak hemispheric preferences (HSP) (Zhou et al. 2020). Furthermore, from our sigmoid, filament and EnHel analysis, eight out of the 22 (36%) sigmoids in the partial match case are observed to not follow the HSP rule (Martin et al. 1994; Rust & Kumar 1996; Liu et al. 2014), and in the full-match case, 2 out of 16 (13%) do not follow this rule indicating a possible relation between the eruptiveness and partial-match between the helicities of different features. This is perhaps because the twists of the emerging fields is opposite to that of the already existing field leading to magnetic reconnections causing eruptions (van Ballegooijen & Martens 1989; Fan & Gibson 2004).

### 4.3. *Suggestive Thresholds for Flaring*

Energy and helicity budgets are important quantities that can be potentially used for determining whether an eruption will occur. We present instantaneous helicity versus energy values for the sigmoids in our study obtained prior to the beginning of their eruptive behaviour in terms of helicity. This is referred to as the Energy-Helicity Diagram and is shown in Fig 5. Tziotziou et al. (2012) and Liokati et al. (2022) performed similar calculations as this work for active regions and found a threshold of $[4 \times 10^{31}$ erg, $2 \times 10^{42}$ $Mx^2]$ and $[5 \times 10^{32}$ erg, $9 \times 10^{41} Mx^2]$, respectively. Our values for the sigmoids compare well with the former with slightly lower thresholds of $[2 \times 10^{31} erg, 1 \times 10^{42} Mx^2]$ for energy and helicity, respectively. The regions that produce most flares fall on the higher ends of the helicity and energy range in the helicity-energy diagram (Fig. 5) and the ones that produce no flares fall on the lower ends.

### 4.4. *Sigmoid Evolution*

We have performed EnHel calculations for one sigmoidal region, that of December 2015 (AR# 12473) that appears like a sigmoid about two days after rising at the east limb to almost until the west limb. Altogether, this region produced 61 flares and 6 CMEs. We observe the changes in flux, helicity and energy along with the force-free factor for the sigmoid, and try to correlate these changes with the occurrences of the eruptive events.

Helicity is thought to be transferred between magnetic structures in the form of twists and braids (Knizhnik et al. 2017) or bulk transfer via the Coriolis force that causes kinking (Liu et al. 2014); smaller field lines in the region can twist among themselves (Pevtsov et al. 2014) and can have opposite signs of helicities compared to the larger overall region. Kusano et al. (2004) suggested via modelling efforts that in regions of sharp reversal of helicities, which may occur due to new emergence of fluxes or due to twisting caused by footpoint motion of the magnetic fields, can get explosively cancelled due to resistive tearing mode instabilities and give rise to solar flares (Yokoyama et al. 2003), thus releasing the stored energy. From Fig. 7 we observe the reduction in the positive helicity values (and consequentially a rise in the alpha parameter towards negative values) during which time an increased number of flares occur. Following this, an increase in both the fluxes is observed suggesting flux-emergence, along with an increase in the positive helicity while the negative helicity remains almost the same, suggesting that the emerging fluxes brought a comparatively greater amount of positive helicity. An increase in the number of flares is also witnessed during this phase. Helicity annihilation (Kusano et al. 2004; Georgoulis et al. 2019) is thought to be the acting mechanism driving these flares. Further analysis using observations in EUV and X-ray at smaller scales near the PILs to verify these claims are warranted and will be partaken in the future.

### 4.5. *Conclusions*

From the results of the first part of this study, we see that the chirality of filaments in our data does not necessarily agree with the handedness of the sigmoids in the corona, suggesting that the helicity sign near the polarity inversion line may be different from that at the outer regions. It is known that both positive and negative helicities are injected into the atmosphere from the convection zone (Gosain & Brandenburg 2019) in the same active region and that the total (relative) helicity is the algebraic sum of these two helicity budgets. However, any region that has dominant helicity (i.e. a significant imbalance between the two signs) usually depicts this dominant helicity via right or left handed curvatures or twists along the field lines, such as the S or Z shaped sigmoids or dextral/sinistral filaments. Active region models developed by Bourdin et al. (2018) to study how helicity changes occur with height suggest that helicity may reverse going from the photosphere into the corona. Singh et al. (2018) used snapshots of coronal flux ropes to show that magnetic helicity reversal occurs within the corona due to the formation of bihelical field lines with opposite signs at different length scales, presumably in different flux systems within the same magnetic structure, given the inverse cascading properties of magnetic helicity (e.g., Berger 1999). We believe that our sigmoid cases with partial match between the calculations from photospheric vector-magnetic fields, Hα filaments and X-ray sigmoid helicity signs, conform to these models and the emergence of opposite helicity structures is possibly the reason for higher eruptive activity compared to the cases with full match. The latter was the case in the study by Chandra et al. (2010) where a complex active region with a dominant negative helicity was observed to have regions with opposite helicity injection, which ultimately caused a filament eruption and a halo-CME that was measured to have a positive helicity magnetic cloud at 1AU. Multi-height observations of filaments and flux ropes at coronal heights using high-resolution instruments from DKIST and Solar Orbiter and their magnetospheric counterparts from Parker Solar Probe (Bale et al. 2016) and the Advanced Composition Explorer (Stone et al. 1998) will be useful in analysing small-scale features and the currents associated with them to better understand the scenarios leading to eruptions in the cases mentioned above and their overall structures.

From the results of the second study wherein sometimes filaments, and sometimes sigmoids, have the same helicity sign as the one obtained from photospheric vector-magnetograms, it is imperative to understand that the two features - filaments and sigmoids are different topological structures. This was also suggested by Régnier & Amari (2004) with an example. The possible reason for the opposite helicities is that the footpoints of the two features are rooted in different sets of emerging magnetic fields which can have different signs of currents and hence helicities (Gibson & Low 2000). The helicities could alternatively be similar to one another. The circumstances under which they may have the same or opposite helicity sign are not clear at this point.

Sigmoids and filaments have mostly been investigated independently, with limited efforts to integrate the two features. It would be beneficial for modelers to undertake more comprehensive analyses that deals with similar and opposite helicity structures in the magnetic environment of the sigmoids to understand these features better.

Furthermore, apart from the topology of filaments and sigmoids, we think it is worth understanding the difference in the helicity and energy budgets for flares and CMEs. The occurrence of flares and CMEs is directly related to the amount of energy and helicity stored in them and each of these phenomena affect Earth's magnetosphere somewhat differently, with flares contributing to ionospheric and impulsive particle acceleration effects and CMEs perturbing the magnetosphere, causing induced currents and particle effects in orbit and on the ground. Flares and CMEs do occur together and knowing the trends in energy, helicity and other magnetic parameters in active regions will help in improving predictive capability of space weather effects. From our results from comparing the sigmoid handedness and filament chirality with each other and with the photospheric values, it is clear that features with opposite helicities existing together are prone to high eruptive activity and must be continuously monitored.

## 5. ACKNOWLEDGEMENTS

We sincerely thank the anonymous referee for their insightful comments and suggestions which have lead to an overall betterment of the manuscript. Hα data were kindly provided by the Kanzelhöhe Observatory (KSO), University of Graz, Austria, and the National Solar Observatory (NSO) Global Oscillations Network Group (GONG). We thank the CDAW team for putting together the CME and Halo−CME data set and providing it online. We acknowledge the usage of SDO/AIA and SDO/HMI−SHARP data. SDO is a mission of the Heliophysics Division's Living With a Star program. We thank the HMI-SHARP team for providing the SHARP data series. SDO data are a courtesy of NASA/SDO and the AIA, EVE, and HMI science teams. We acknowledge using the XRT data from Hinode. Hinode is a Japanese mission developed and launched by ISAS/JAXA, with NAOJ as domestic partner and NASA and STFC (UK) as international partners. It is operated by these agencies in co-operation with ESA and the NSC (Norway). V.A. acknowledges support from the NASA FINESST grant number 80NSSC19K1437. The presented work is part of V.A's dissertation (Aparna 2022). V.A. acknowledges Georgia State University for their kind support while conducting the dissertation research there.

**Table 4.** Helicity signs of filaments, sigmoids and EhHel calculations.

| # | Date | Sigmoid Shape | Helicity Sign | Filament Chirality | Helicity Sign | Alpha α/Mm | Std. Dev. | Uncertainty δ α | No. of Turns | Flux Balance % | Flare Index | Location on the Sun |
|---|---|---|---|---|---|---|---|---|---|---|---|---|
| 1 | 05/04/10 | Z | − | R | − | −0.09 | 0.02 | 0.007 | 6.69 | 9.2 | 0.000 | S27W09 |
| 2 | 07/15/10 | S | + | L | + | −0.06 | 0.04 | 0.009 | 3.3 | 6.11 | 0.000 | S22W10 |
| 3 | 08/04/10 | Z | − | R | − | −0.004 | 0.01 | 0.01 | 0.8 | 0.72 | 0.000 | N15W18 |
| 4 | 08/07/10 | Z | − | R | − | 0.03 | 0.006 | 0.02 | 2.7 | 11.8 | 0.000 | N12E31 |
| 5 | 12/10/10 | Z | − | R | | −0.02 | 0.01 | 0.02 | 3.7 | 0.61 | 0.329 | N15W09 |
| 6 | 04/10/11 | S | + | A | | 0.02 | 0.01 | 0.02 | 1.7 | 1.57 | 0.000 | S18E24 |
| 7 | 9/13/11 | S | − | L | + | −0.0056 | 0.0074 | 0.03 | 1.3 | 6.21 | 0.729 | N22W14 |
| 8 | 11/24/11 | Z | | L | + | 0.01 | 0.04 | 0.046 | 2.7 | 3.6 | 0.000 | N14W08 |
| 9 | 05/05/12 | S | + | L | + | 0.04 | 0.04 | 0.059 | 6.6 | 3.65 | 3.900 | S19W24 |
| 10 | 05/08/12 | Z | | L | + | −0.031 | 0.01 | 0.041 | 3.4 | 6.53 | 0.000 | N13E03 |
| 11 | 06/15/12 | Z | | L | + | −0.01 | 0.08 | 0.077 | 2.1 | 5.6 | 2.757 | S16W13 |
| 12 | 07/01/12 | Z | | R | | −0.032 | 0.05 | 0.083 | 5.4 | 0.88 | 1.967 | S15E01 |
| 13 | 07/12/12 | S | + | R | | −0.016 | 0.05 | 0.067 | 2.1 | 3.9 | 20.086 | S17W08 |
| 14 | 10/21/12 | S | | R | | −0.028 | 0.02 | 0.043 | 3.8 | 7.83 | 1.000 | S27W17 |
| 15 | 11/09/12 | S | + | L | + | 0.012 | 0.03 | 0.04 | 1.6 | 9.43 | 1.000 | S20E09 |
| 16 | 11/30/12 | Z | | R | | −0.02 | 0.02 | 0.032 | 3.3 | 9.01 | 0.557 | N15W04 |
| 17 | 12/01/10 | Z | | R | | −0.02 | 0.02 | 0.032 | 3.2 | 9.01 | 0.000 | No AR |
| 18 | 01/29/13 | S | + | R | − | 0.06 | 0.03 | 0.01 | 5.5 | 10.6 | 0.183 | S11E05 |
| 19 | 07/01/13 | S | + | L | + | 0.04 | 0.02 | 0.02 | 4.3 | 2.2 | 0.000 | S15E04 |
| 20 | 07/16/13 | Z | − | N | | 0.03 | 0.05 | 0.01 | 2.04 | 8.7 | 0.000 | S14W10 |
| 21 | 08/08/13 | Z | | R | − | −0.1 | 0.04 | 0.01 | 9.05 | 3.8 | 2.340 | N13W33 |
| 22 | 10/16/13 | S | + | L | + | 0.04 | 0.03 | 0.009 | 4.5 | 6.05 | 0.800 | S21W35 |
| 23 | 11/17/13 | S | + | R | − | −0.03 | 0.05 | 0.004 | 7.35 | 12.25 | 5.000 | S18W54 |
| 24 | 11/20/13 | Z | − | L | − | 0.04 | 0.05 | 0.01 | 12.5 | 10.8 | 0.000 | N05W29 |
| 25 | 01/15/14 | S | + | L | + | 0.01 | 0.04 | 0.01 | 1.14 | 3.1 | 0.000 | S16W20 |

**Table 4** (*continued*)

| # | Date | Sigmoid Shape | Sigmoid Helicity Sign | Filament Chirality | Filament Helicity Sign | Alpha α/Mm | Std. Dev. | Uncertainty δ α | No. of Turns | Flux Balance % | Flare Index | Location on the Sun |
|---|---|---|---|---|---|---|---|---|---|---|---|---|
| 26 | 03/26/14 | S | + | R | − | 0.002 | 0.04 | 0.01 | 0.18 | 3.6 | 29.329 | N10E08 |
| 27 | 05/01/14 | Z | − | L | + | 0.01 | 0.04 | 0.01 | 1.4 | 5.8 | 0.000 | N12E17 |
| 28 | 6/24/14 | S | + | L | + | 0.03 | 0.03 | 0.01 | 3.65 | 15.1 | 0.000 | S18W37 |
| 29 | 10/02/14 | Z | − | R | − | −0.02 | 0.06 | 0.01 | 2.1 | 13.4 | 0.000 | N12W11 |
| 30 | 11/03/14 | Z | − | L | + | 0.07 | 2.34 | 0.01 | 5.45 | 7.6 | 0.000 | N12W35 |
| 31 | 11/12/14 | S | + | L | + | −0.01 | 0.05 | 0.006 | 2.36 | 17.16 | 1.840 | N15W35 |
| 32 | 12/15/14 | Z | − | L | + | 0.04 | 0.04 | 0.02 | 7.42 | 15.3 | 0.260 | N09W30 |
| 33 | 02/22/15 | Z | − | S | | −0.03 | 0.05 | 0.01 | 2.83 | 17.1 | 0.000 | N14W31 |
| 34 | 09/01/15 | Z | − | L | + | −0.05 | 0.02 | 0.02 | 3.63 | 3.87 | 0.000 | N05W05 |
| 35 | 10/21/15 | S | + | R | − | −0.002 | 0.04 | 0.007 | 0.69 | 2.89 | 1.720 | S10W36 |
| 36 | 11/5/15 | S | + | R | − | 0.014 | 0.04 | 0.005 | 2.37 | 4.1 | 1.200 | N06W22 |
| 37 | 01/10/16 | S | + | R | − | 0.015 | 0.03 | 0.005 | 1.45 | 9.94 | 0.733 | N03E19 |
| 38 | 08/28/16 | Z | − | R | − | 0.03 | 0.04 | 0.01 | 2.64 | 9.79 | 0.000 | N12W06 |
| 39 | 09/19/16 | Z | − | N | | 0.05 | 0.01 | 0.01 | 3.67 | 1.78 | 0.000 | N14E15 |
| 40 | 10/8/16 | S | + | R | − | 0.01 | 0.04 | 0.004 | 1.1 | 7.52 | 0.138 | S15W08 |
| 41 | 01/22/17 | S | + | R | | −0.029 | 0.03 | 0.045 | 2 | 7.8 | 0.400 | N06E20 |
| 42 | 07/11/17 | Z | − | L | + | 0.015 | 0.05 | 0.003 | 2.69 | 4.62 | 8.186 | S06W03 |
| 43 | 07/23/17 | Z | − | L | + | 0.05 | 0.05 | 0.03 | 4.24 | 0.86 | 0.000 | N04E05 |
| 44 | 08/20/17 | Z | − | R | − | −0.03 | 0.08 | 0.004 | 6.05 | 16.13 | 2.775 | N12W06 |
| 45 | 09/26/17 | S | + | L | | −0.0006 | 0.024 | 0.012 | 0.08 | 1.16 | 0.000 | S12W10 |
| 46 | 10/29/17 | Z | − | R | − | −0.01 | 0.027 | 0.025 | 1.65 | 13.9 | 0.000 | N13W10 |